\documentclass[fleqn,usenatbib]{mnras}

\usepackage{newtxtext,newtxmath}

\usepackage[T1]{fontenc}
\usepackage{ae,aecompl}

\usepackage{graphicx}	
\usepackage{changes}
\colorlet{Changes@Color}{red}

\title[Orbital Parameter Distribution of Merging Halos]{The Origin of the Orbital Parameter Distribution of Merging Halos}
\author[A. J. Benson]{Andrew J. Benson\,$^{1}$\thanks{E-mail: abenson@carnegiescience.edu}\\
$^{1}$ Carnegie Observatories, 813 Santa Barbara Street, Pasadena, CA 91101, USA\\
}

\begin{document}

\maketitle

\begin{abstract}
 We describe a simple model which explains the qualitative and (approximate) quantitative features of the distribution of orbital velocities of merging pairs of dark matter halos. Our model considers a primary dark matter halo as a perturber in a background of secondary halos with velocities described by linear theory. By evaluating the ensemble of secondary halos on orbits within the perturbing halo's ``loss cone'' we derive the distribution of orbital parameters of these captured halos. This model is able provide qualitative explanations for the features of this distribution as measured from N-body simulations, and is in approximate quantitative agreement with those measurements. As the velocity dispersion of the background halos is larger on smaller scales our model predicts an overall increase in the characteristic velocities of merging halos, relative to the virial velocities of those halos, in lower mass systems. Our model also provides a simple explanation for the measured independence of the orbital velocity distribution function on redshift when considered at fixed peak height. By connecting the orbital parameter distribution to the underlying power spectrum our model also allows for estimates to be made of the effect of modifying that power spectrum, for example by including a truncation at large wavenumber. For plausible warm dark matter models we find that this truncation has only a small effect on the predicted distributions.
\end{abstract}


\section{Introduction}

Understanding the dark matter subhalo population---in particular its distributions in mass, host halo-centric radius, etc.--- is of key importance for many studies aimed at constraining the nature of dark matter \citep{gilman_warm_2020,bose_little_2020,nadler_dark_2021}. While N-body and hydrodynamical simulations allow a direct calculation of these distributions, the computational cost of generating large numbers of high resolution realizations (needed to quantify statistical properties) remains high. Semi-analytic models for the subhalo population \citep{taylor_dynamics_2001,benson_effects_2002,zentner_physics_2005,pullen_nonlinear_2014,yang_new_2020,jiang_satgen_2020}, while being more approximate, achieve much lower computational costs. These models solve coupled sets of differential equations which describe the orbital position and velocity, the bound mass of the subhalo, and its density profile.

A key input to these models are the initial conditions for these differential equations. Typically these are taken to be the orbital parameters of the subhalo at the point where it first crosses the virial radius of its host. Under the usual assumption of an isotropic distribution of merging halos, these initial conditions are fully specified by the combination of the radial and tangential velocity of the subhalo (or any two equivalent quantities, such as the energy and angular momentum).

Distributions of orbital velocities for merging subhalos have been measured by several authors \citep{benson_orbital_2005,wetzel_orbits_2011,jiang_orbital_2015,li_orbital_2020}. In these works, the distributions were obtained by identifying merging halo pairs in cosmological N-body simulations of cold dark matter, and constructing 2D histograms of their orbital velocities at the virial radius. These works show that radial and tangential velocities are strongly correlated---with the distribution peaking along the ridge line of constant total velocity \citep{jiang_orbital_2015}---with mean values slightly lower than the virial velocity of the host, and with only weak dependence on the host mass, mass ratio, and redshift (when expressed in virial units).

The goal of this work is to gain some insight into the physics which sets the distribution of orbital parameters of merging dark matter halos. Our motivation is to provide both some understanding of the physical origin of this distribution, and a simple model for how this distribution may be expected to respond to changes in the matter power spectrum (such as might arise in non-cold dark matter models). Given that, by definition, mergers between halos happen far into the non-linear regime of gravitational structure formation, we do not expect our results to necessarily be quantitatively precise.

Our approach is to consider the more massive (``primary'') halo embedded in a background of ``secondary'' halos which have a cosmological distribution of velocities, and to consider which of those halos will be gravitationally captured by the primary, i.e. those whose orbits have pericentric distances that lie within the virial radius of the primary---similar to ``loss cone'' calculations employed for supermassive black holes \citep{merritt_loss-cone_2013}. Evaluating the velocities of these captured secondary halos at the primary halo virial radius, and integrating over all possible such halos leads to a model of the orbital parameter distribution function.

The remainder of this paper is organized as follows. In \S\ref{sec:methods} we describe our method for computing the orbital parameter distribution function. In \S\ref{sec:results} we show results from this method, and compare them to previous determinations of the distribution function. We also explore how the distribution function is expected to change in the low mass halo regime, examine any evolution with redshift, and explore the effects of a cut-off in the matter power spectrum. Finally, in \S\ref{sec:discussion} we discuss the outcomes and implications of our work.

Throughout this work will will adopt a cosmological model described by $(H_0,\Omega_\mathrm{m},\Omega_\Lambda,\Omega_\mathrm{b})=(67.66\hbox{km/s},0.307,0.693,0.0482)$ and a matter power spectrum characterised by $(\sigma_8,n_\mathrm{s})=(0.8228,0.96)$ \citep{planck_collaboration_planck_2014} although our model applies equally well to other cases.

\section{Methods}\label{sec:methods}

In the following we refer to the perturber halo as the ``primary'' halo, with any halo from the background population referred to as the ``secondary'' halo. Properties of these halos are identified using subscripts ``p'' and ``s'' respectively. Without any loss of generality we consider $M_\mathrm{p,v} \ge M_\mathrm{s,v}$, where $M_\mathrm{v}$ is the virial mass of the halo, defined as the mass within a sphere enclosing a mean density contrast of $\Delta_\mathrm{v}$ as computed under the standard spherical collapse model \citep[e.g.][]{percival_cosmological_2005}.

Throughout this work we work in virial units, such that all lengths are measured in units of the primary halo virial radius, $r_\mathrm{p,v}$, and all velocities in units of the primary halo virial velocity, $V_\mathrm{p,v}=(\mathrm{G}M_\mathrm{p,v}/r_\mathrm{p,v})^{1/2}$.

We assume that the distribution of orbital parameters is given by\footnote{We could choose to integrate over any of the three variables, $r$, $V^\prime_\mathrm{r}$, or $V^\prime_\theta$ in this equation, providing the other two variables were appropriately related to the variable integrated over, and the appropriate Jacobian used in place of the one used here. While integrating over $r$ would lead to a more compact expression for the Jacobian we find that integrating over $V^\prime_\mathrm{r}$ is computationally more robust and efficient.}
\begin{equation}
 p(V_\mathrm{r},V_\theta) \propto |V_\mathrm{r}| \int_{-\infty}^{+\infty} \mathrm{d}V^\prime_\mathrm{r} 4 \pi r^2 [1+\xi_\mathrm{ps}(r)] p(V^\prime_\mathrm{r},V^\prime_\theta|r) |\mathbfss{J}|,
 \label{eq:distribution}
\end{equation}
where $(V_\mathrm{r},V_\theta)$ are the radial and tangential velocity of the orbit at the primary halo virial radius, $(V^\prime_\mathrm{r},V^\prime_\theta)$ are the corresponding velocities at a radius $r$ from the primary halo, $p(V^\prime_\mathrm{r},V^\prime_\theta|r)$ is the distribution function of secondary halos in velocity at radius $r$, $\xi_\mathrm{ps}(r)$ is the two point correlation function of primary and secondary halos, and $\mathbfss{J}$ is the Jacobian of the transformation from coordinates $\mathbf{x}^\prime=(r,V^\prime_\theta)$ to $\mathbf{x}=(V_\mathrm{r},V_\theta)$, for which the determinant is
\begin{equation}
 |\mathbfss{J}| =\pm \frac{V_\mathrm{r}}{-V_\mathrm{r}^2+V^{\prime 2}_\mathrm{r}-V_\theta^2+2} \left(\frac{1}{\sqrt{V_\theta^2 \left(V_\mathrm{r}^2-V^{\prime 2}_\mathrm{r}-2\right)+V_\theta^4 + 1}} \pm 1\right),
\end{equation}
with the ``$\pm$'' corresponding to the two possible solutions for the radius, $r$.

Note that the initial factor of $|V_\mathrm{r}|$ in equation~(\ref{eq:distribution}) arises from the fact that secondary halos cross the virial radius of the primary halo (i.e. they merge with it) at a rate proportional to their radial velocity at the virial radius \citep{benson_orbital_2005}. Note also that we include contributions from halos which are initially ingoing (i.e. $V^\prime_\mathrm{r} < 0$ and outgoing $V^\prime_\mathrm{r} > 0$).

The velocity $(V^\prime_\mathrm{r},V^\prime_\theta)$ is computed from $(V_\mathrm{r},V_\theta)$ and $r$ under the assumption that the secondary halo moves in a Keplerian potential\footnote{We discuss the validity of this assumption in \S\protect\ref{sec:discussion}.} corresponding to the mass of the primary halo. We exclude from the integral radii $r < 1$ (i.e. within the virial radius of the primary halo), and those for which the time-of-flight from $r$ to the primary halo virial radius exceeds the age of the Universe (this latter condition has little effect on the results).

Evaluating the above distribution  then only requires models for $\xi_\mathrm{ps}(r)$ and $p(V^\prime_\mathrm{r},V^\prime_\theta|r)$. We assume that $\xi_\mathrm{ps}(r)=b(M_\mathrm{p})b(M_\mathrm{s})\xi_\mathrm{lin}(r)$, where $b(M)$ is the bias of halos of mass $M$, and $\xi_\mathrm{lin}(r)$ is the linear theory two-point correlation function, which is computed via the Fourier transform of the power spectrum. For the bias, $b(M)$, we use the model of \cite{tinker_large_2010}.

We assume that the distribution, $p(V^\prime_\mathrm{r},V^\prime_\theta|r)$, can be modelled by assuming that the relative velocity of primary and secondary halos (prior to any perturbation arising from the primary halo) is described by uncorrelated Gaussian distributions with the same dispersion in each direction \citep[e.g.][]{sheth_peculiar_2001}, and therefore that $p^\prime(V^\prime_\mathrm{r},V^\prime_\theta|r)$ is a product of a Gaussian distribution in the radial direction and a Rayleigh distribution in the tangential direction,
\begin{equation}
 p^\prime(V^\prime_\mathrm{r},V^\prime_\theta|r) \propto \frac{1}{\sigma^3(r)} \exp\left( -\frac{[V^\prime_\mathrm{r}-\bar{V}^\prime_\mathrm{r}(r)]^2}{2\sigma^2(r)} \right) V^\prime_\theta \exp\left(-\frac{V^{\prime2}_\theta}{2\sigma^2(r)}\right),
\end{equation}
where we assume that there may be some non-zero mean relative radial velocity, $\bar{V}^\prime_\mathrm{r}(r)$.

It is well-established that the cosmological pairwise velocity distribution function is, in fact, non-Gaussian, showing extended tails and skewness \citep{sheth_peculiar_2001,scoccimarro_redshift-space_2004} due to non-linear contributions to the pairwise velocities of halos arising from scales much smaller than the halo separation. The core of the distribution is generally well represented by a Gaussian \cite[e.g.][]{cuesta-lazaro_towards_2020}, with the non-Gaussian components affecting only the tails. Therefore, for simplicity, we do not consider these non-Gaussian features of the pairwise velocity distribution here.

For $\bar{V}^\prime_\mathrm{r}(r)$ and $\sigma(r)$ we adopt the predictions from linear perturbation theory \citep[e.g.][eqn.~15]{sheth_linear_2001}:
\begin{equation}
 \bar{V}^\prime_\mathrm{r}(r) = -H r \left( 1 + \frac{2}{3} \frac{f(a) \bar{\xi}(r)}{1+\xi(r)} \right),
\end{equation}
where $\bar{\xi}(r)$ is the volume-averaged correlation function, and $f(a)=\mathrm{d}\log D / \mathrm{d} \log a$ is the growth rate of the linear growth factor, $D(a)$, and we have included the Hubble expansion, and \citep[e.g.][eqn.~30, with the virial terms set to zero]{sheth_linear_2001}
\begin{equation}
 \sigma^2(M) = \alpha \left[ \sigma^2_\mathrm{h}(M_\mathrm{p}) + \sigma^2_\mathrm{h}(M_\mathrm{s}) - 2 \Psi(M_\mathrm{p},M_\mathrm{s}|r) \right],
 \label{eq:dispersion}
\end{equation}
evaluated at the Lagrangian radius of each mass shell, where $\alpha$ is a parameter (expected to be of order unity) which we introduce to allow some freedom in matching results from N-body simulations, and where \citep[][eqn.~8]{sheth_peculiar_2001}
\begin{equation}
 \sigma_\mathrm{h}(r) = \frac{1}{\sqrt{3}} \sigma_{-1}(M) C(M),
 \label{eq:sigmaj}
\end{equation}
with \citep[][unnumbered equation after eqn.~8]{sheth_peculiar_2001}
\begin{equation}
 \sigma_j^2(M) = \frac{H^2 f^2(a)}{2 \pi^2} \int \mathrm{d} k k^{2+2j} P(k) W^2(k|M),
\end{equation}
being the linear theory prediction for the velocity dispersion when filtered on a scale corresponding to mass $M$, and $W(k|M)$ is the Fourier transform of a top-hat window function. Note that we have included a factor of $\sqrt{3}$ in equation~(\ref{eq:sigmaj}) since we want the 1D velocity dispersion. The factor
\begin{equation}
 C(M) = \sqrt{1-\sigma_0^4/\sigma_1^2\sigma_{-1}^2}
\end{equation}
\citep[][eqn.~8]{sheth_peculiar_2001} accounts for the fact that halos form at special locations in the density field (i.e. peaks), and where \citep[][eqn.~29]{sheth_linear_2001}
\begin{equation}
 \Psi(M_\mathrm{p},)M_\mathrm{s}|r) = C(M_\mathrm{p}) C(M_\mathrm{s}) \psi(M_\mathrm{p},M_\mathrm{s}|r),
\end{equation}
with \citep[][eqn.~28]{sheth_linear_2001}
\begin{equation}
 \psi(M_\mathrm{p},)M_\mathrm{s}|r) = H^2 f^2(a) \int \frac{\mathrm{d} k}{2\pi^2} P(k) W(k|M_\mathrm{p}) W(k|M_\mathrm{s}) K(kr),
\end{equation}
and \citep[][unnumbered equation after eqn.~28]{sheth_linear_2001}
\begin{equation}
 K(x) = \frac{\sin x}{x} - \frac{2}{x^3} (\sin x - x \cos x).
\end{equation}
We estimate the Lagrangian radius of each mass shell as $r_\mathrm{L} = [\Delta_\mathrm{v} + (1+\bar{\xi}(r)) r^3]^{1/3}$ where $r$ is the current, Eulerian radius of the shell.

As demonstrated by \cite{sheth_peculiar_2001}, the velocity dispersion has a significant dependence on environment. Qualitatively, denser regions of the universe evolve more rapidly and so velocities grow more quickly in such regions. Since halos can also preferentially populate higher density regions \citep[e.g.][]{mo_analytic_1996}, and halo merger rates increase as a function of environmental density \citep{fakhouri_environmental_2009}, we must account for this environmental dependence in our model. To approximately account for this effect we use the result of \cite{sheth_peculiar_2001} that $\sigma(r,\delta_\mathrm{nl,e}) = (1+\delta_\mathrm{nl,e})^{\mu(r_\mathrm{e})} \sigma(r)$ where $\sigma(\delta_\mathrm{nl,e})$ is the velocity dispersion in regions of nonlinear environmental overdensity $\delta_\mathrm{nl,e}$, $\sigma(r)$ is the dispersion computed using linear theory (as described above), and $\mu(r_\mathrm{e}) \approx 0.6 S(r_\mathrm{e})/S(10h^{-1}\hbox{Mpc})$, where $S(r)$ is the linear theory variance of the density field when smoothed on a scale $r$, was found by fitting their measurements from N-body simulations. We then compute a mean boost in the velocity dispersion as
\begin{eqnarray}
 \beta(M_\mathrm{p},r) &=& \int_{-\infty}^{+\infty} \mathrm{d} \delta_\mathrm{e} \, p(\delta_\mathrm{e}) n(M_\mathrm{p},\delta_\mathrm{e}) R(M_\mathrm{p},\delta_\mathrm{nl,e}) (1+\delta_\mathrm{nl,e})^{\mu(r_\mathrm{e})} \nonumber \\
 & & \left/  \int_{-\infty}^{+\infty} \mathrm{d} \delta_\mathrm{e} \, p(\delta_\mathrm{e}) n(M_\mathrm{p},\delta_\mathrm{e}) R(M_\mathrm{p},\delta_\mathrm{nl,e}), \right.
 \label{eq:environment}
\end{eqnarray}
where $\delta_\mathrm{e}$ is the linear theory overdensity of the environment, $p(\delta_\mathrm{e})$ is the distribution function of this overdensity, $n(M,\delta_\mathrm{e})$ is the environment-dependent halo mass function, and $R(M,\delta_\mathrm{nl,e})$ is the factor by which the merger rate of halos is enhanced as a function of environmental density. We define our environment on a scale of 20~Mpc, and assume $p(\delta)$ is described by a normal distribution with root-variance $\sqrt{S}(R=20\hbox{Mpc})$ conditioned on the fact that the region has not collapsed to become a halo on any larger scale. We compute the mass function using the peak-background split approach \citep{bond_excursion_1991,bond_peak-patch_1996} applied to the Sheth-Tormen mass function, and for $R(M_\mathrm{p},\delta_\mathrm{e})$ use the dependence on environment measured by \cite{fakhouri_environmental_2009} from N-body simulations. This boost factor is close to 1 for low mass halos, but reaches approximately 3 for halos of mass $10^{15}\mathrm{M}_\odot$. 

Using the above, the orbital parameter distribution at the virial radius can be evaluated for any combination of primary and secondary halos.

\section{Results}\label{sec:results}

\begin{figure*}
 \begin{tabular}{cc}
  \includegraphics[width=85mm]{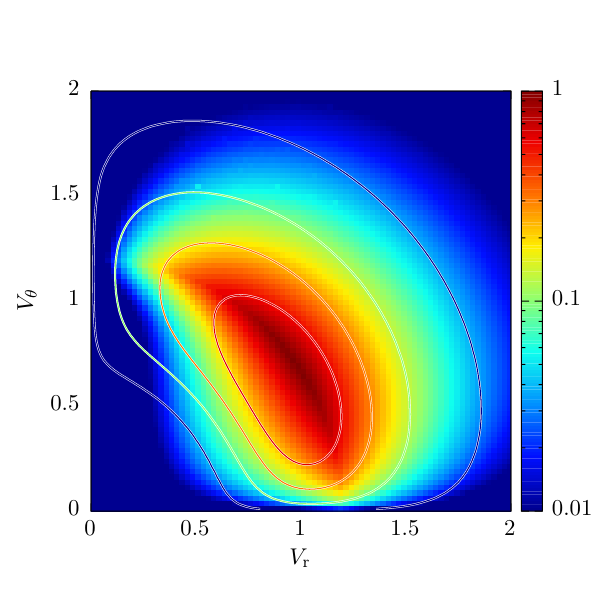} &
  \includegraphics[width=75mm]{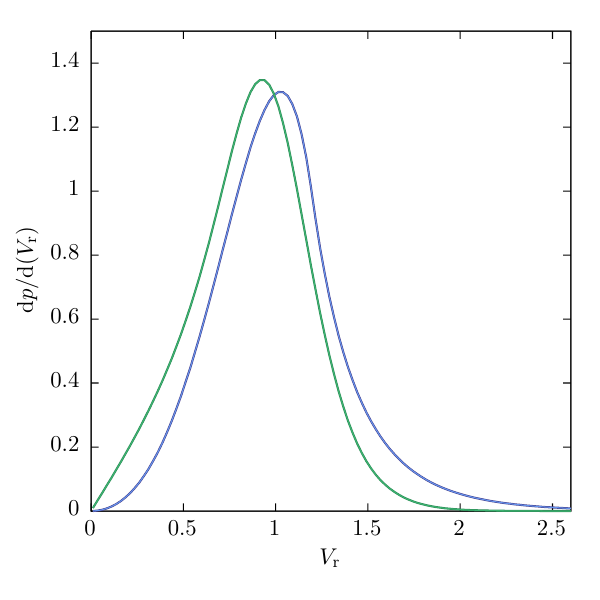} \\
  \includegraphics[width=75mm]{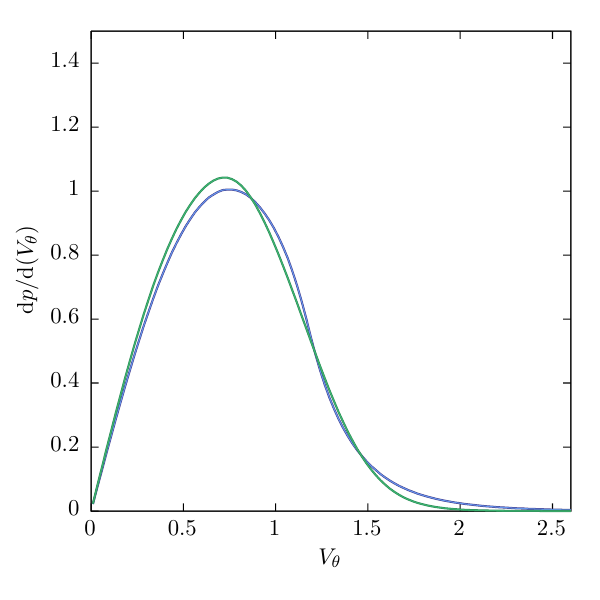} &
  \includegraphics[width=75mm]{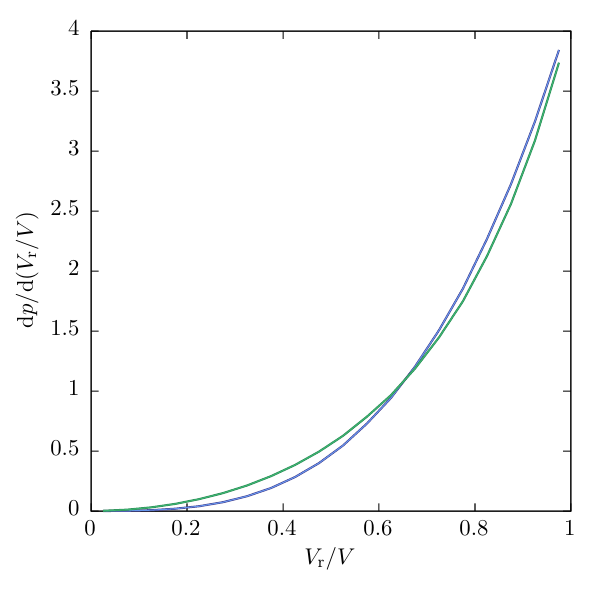}
 \end{tabular}
 \caption{Distributions of orbital parameters for merging halos of masses $M_\mathrm{p}=10^{12}\mathrm{M}_\odot$ and $M_\mathrm{s}=10^{11}\mathrm{M}_\odot$ at $z=0$. \emph{Upper left panel:} The joint distribution of radial and tangential velocities of merging halos. The colour map shows results from our model, while the contours indicate the distribution of \protect\cite{li_orbital_2020}. \emph{Upper right panel:} The distribution of radial velocities of merging halos. The blue curve shows results from our model, while the green line is the distribution of \protect\cite{li_orbital_2020}. \emph{Lower left panel:} As the upper right panel, but for the tangential velocity. \emph{Lower right panel:} As the upper right panel, but for the radial velocity expressed as a fraction of the total velocity.}
 \label{fig:distributions}
\end{figure*}

We use our model to compute the distribution of $(V_\mathrm{r},V_\theta)$ for various combinations of primary and secondary halo mass at $z=0$. The upper-left panel of Figure~\ref{fig:distributions} shows the resulting distribution function for $M_\mathrm{p}=10^{12}\mathrm{M}_\odot$ and $M_\mathrm{s}=10^{11}\mathrm{M}_\odot$. The rainbow colour scale shows the results of our model, while contours show the model of \cite{li_orbital_2020} at levels of 0.01, 0.10, 0.30, and 0.60 for comparison. The upper-right and lower-left panels show the marginal distributions of $V_\mathrm{r}$ and $V_\theta$ respectively, with results from our model shown by the blue lines, and those of \cite{li_orbital_2020} by the green lines. Finally, the lower-right panel shows the distribution of $V_\mathrm{r}/V$ where $V=(V_\mathrm{r}^2+V_\theta^2)^{1/2}$, again with results from our model shown by the blue lines, and those of \cite{li_orbital_2020} by the green lines.

We find that a value of $\alpha = 0.8$ in equation~(\ref{eq:dispersion}) gives a reasonable match to the N-body results from \cite{li_orbital_2020}. Had we instead taken $\alpha=1$ the predicted distributions of $V_\mathrm{r}$ and $V_\theta$ would be slightly broader, and the peak of the $V_\theta$ distribution shifted to slightly higher velocity, but the location of peak of the $V_\mathrm{r}$ distribution would be largely unchanged. Given the simplifying assumptions of our model a value of $\alpha = 0.8$ seems acceptable. There is good qualitative agreement in the locus of our distribution function with that of \cite{li_orbital_2020}, although in detail they do not agree. There is also good agreement in the marginal distributions. For $V_\mathrm{r}$ our model predicts approximately the correct shape and width of the distribution, although it is shifted to slightly too large values. For $V_\theta$ there is excellent agreement between our model and the results of \cite{li_orbital_2020} in both the peak location, width of the distribution, and overall shape. Similarly, our model accurately matches the form of the distribution of $V_\mathrm{r}/V$ found by \cite{li_orbital_2020}. In Appendix~\ref{sec:appendix} we show similar results for $M_\mathrm{p}=10^{14}\mathrm{M}_\odot$ and $M_\mathrm{s}=10^{11}\mathrm{M}_\odot$ where our model performs less well, due to underestimating the dispersion of the secondary halo population.

Given this degree of success in matching measurements of the orbital parameter distribution in N-body simulations, we can utilize our model to provide some insights into the properties of the distribution. For example, the ridge of the distribution of $(V_\mathrm{r},V_\theta)$ approximately tracks a line of constant $V^2=V_\mathrm{r}^2+V_\theta^2 \equiv V_\mathrm{l} \approx 1.5$, as was first noted by \cite{jiang_orbital_2015}. This therefore corresponds to a locus of constant energy (since all orbits are measured at the same radius, the primary halo virial radius, and so at the same potential). 

Orbits with infall velocities, $V > V_\mathrm{l}$, in excess of the velocity of this locus are either unbound, or close to unbound. The radial velocity of such orbits remains non-zero out to large distances, unlike orbits with $V < V_\mathrm{l}$ which reach zero radial velocity (i.e. apocenter, $r_\mathrm{apo}$) at radii of 3--4 virial radii or less. As such, the contribution of these orbits is strongly down-weighted by the distribution function, $p^\prime(V^\prime_\mathrm{r},V^\prime_\theta)$, at the evaluation radius as $V^\prime_\mathrm{r}$ is significantly offset from $\bar{V}^\prime_\mathrm{r}$. (For very large radii this effect is further enhanced by the Hubble expansion.) This is the cause in the decline in the distribution for $V > V_\mathrm{l}$.

For $V < V_\mathrm{l}$ the decline in the distribution function is driven by the available volume. These orbits have apocenters within at most a few virial radii, with the apocentric radius decreasing as $V$ decreases. As we evaluate the contribution of each orbit only over the radial range $1(\equiv r_\mathrm{v})$ to $r_\mathrm{apo}$ there is little volume contributing to these regions of the distribution function.

The overall width of the distribution is controlled by the velocity dispersion of the background halos (and, therefore, is fundamentally determined by the matter power spectrum). Without the smoothing effect of the velocity dispersion, the distribution would have a much sharper cut off beyond $V_\mathrm{l}$, and similar (but somewhat less sharp) cut off for $V < V_\mathrm{l}$. Artificially increasing/decreasing $\sigma(r)$ results in a corresponding increase/decrease in the width of the distributions of $V_\mathrm{r}$ and $V_\theta$, along with a positive/negative shift in the location of the peak of the $V_\theta$ distribution. The location of the peak in the $V_\mathrm{r}$ distribution is largely unaffected by such changes in $\sigma(r)$---it is instead mostly determined by the arguments given above for the location of the ridge line in the $(V_\mathrm{r},V_\theta)$ distribution.

\begin{figure*}
 \begin{tabular}{cc}
  \includegraphics[width=85mm]{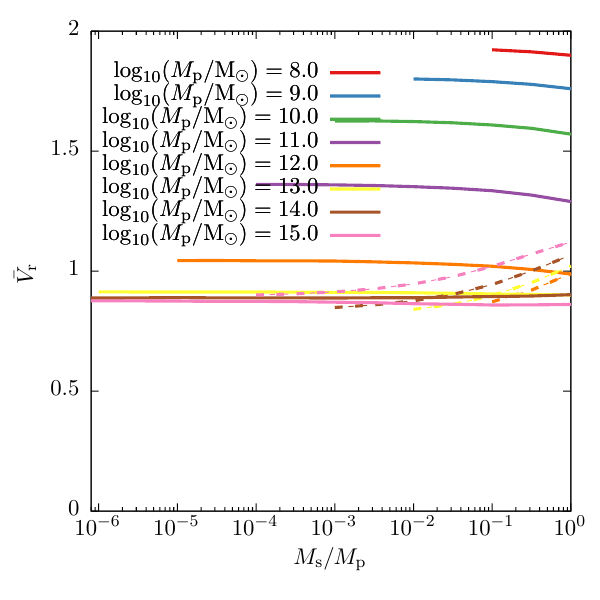} &
  \includegraphics[width=85mm]{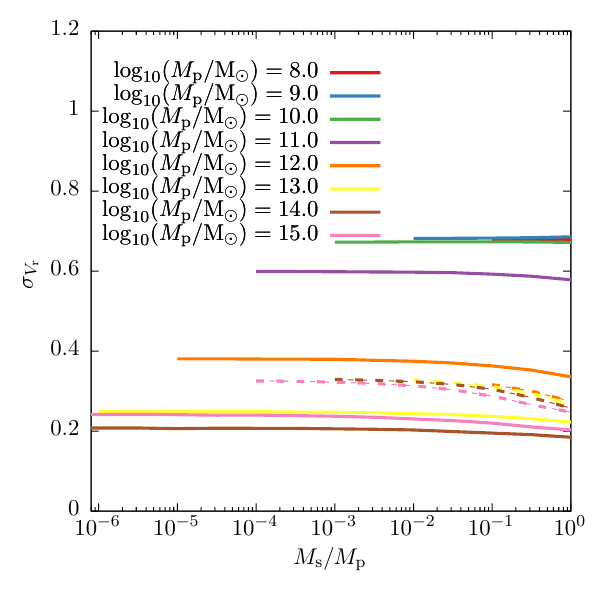} \\
  \includegraphics[width=85mm]{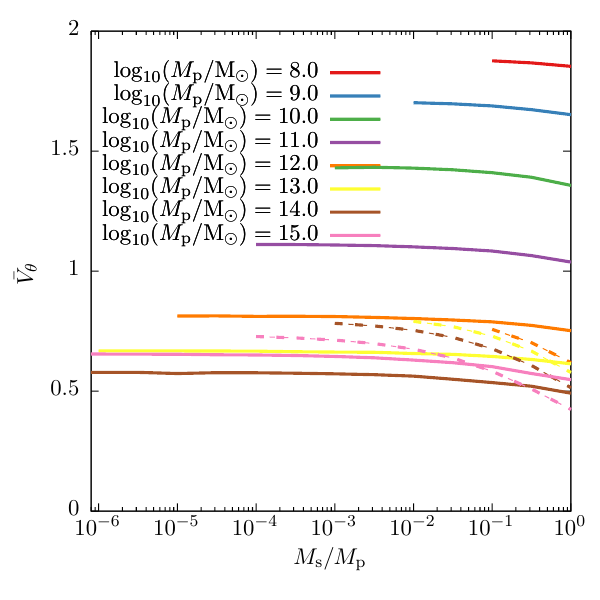} &
  \includegraphics[width=85mm]{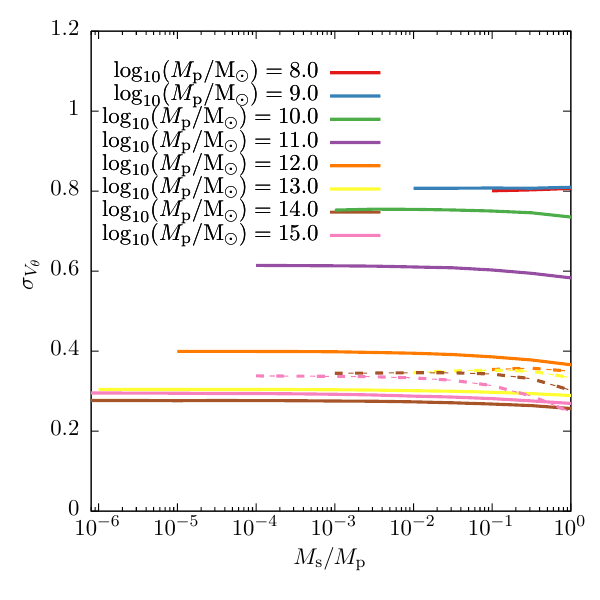}
 \end{tabular}
 \caption{Summary statistics of the orbital velocity distribution function as predicted by our model (solid lines), and as measured by \protect\cite[dashed lines]{li_orbital_2020}. For the \protect\cite{li_orbital_2020} results we plot lines only over the range of halo masses considered in that work. Statistics shown are the mean ($\bar{V}$; left panels), and dispersion ($\sigma$; right panels), of the radial (upper panels) and tangential (lower panels) orbital velocities), as a function of the secondary-to-primary halo mass ratio, $M_\mathrm{s}/M_\mathrm{p}$. Lines colours indicate different primary halo masses, as indicated in each panel.}
 \label{fig:summaryStats}
\end{figure*}

In Figure~\ref{fig:summaryStats} we show summary statistics (mean, $\bar{V}$, and dispersion, $\sigma$, of the radial and tangential orbital velocities) as predicted by our model (solid lines), and as measured by \citeauthor{li_orbital_2020}~(\citeyear{li_orbital_2020}; dashed lines). For the \protect\cite{li_orbital_2020} results we plot lines only over the range of halo masses considered in that work. Over the range of halo masses where \cite{li_orbital_2020} measured the orbital velocity distribution our model is in reasonable agreement with their results. Notably, over the range of primary halo masses considered by \cite{li_orbital_2020} our model predicts a weak dependence of these summary statistics on $M_\mathrm{p}$, although not as weak as found by \cite{li_orbital_2020}. \cite{li_orbital_2020} also find that these summary statistics are independent of the primary--secondary halo mass ratio for $M_\mathrm{s}/M_\mathrm{p} < 10^{-2}$, with a weak dependence at higher mass ratios. Our model similarly predicts no dependence on mass ratio for $M_\mathrm{s}/M_\mathrm{p} < 10^{-2}$. At higher mass ratios our model predicts some dependence, but much weaker than that found by \cite{li_orbital_2020}.

\begin{figure}
\includegraphics[width=85mm]{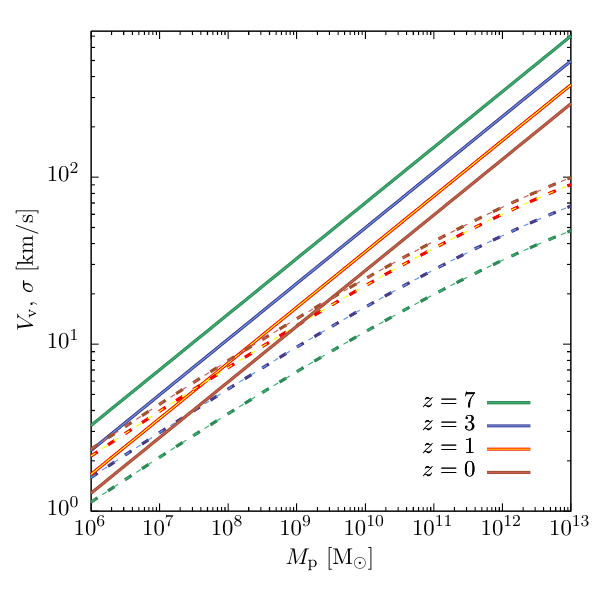}
\caption{Virial velocities (solid lines) and halo-halo pairwise velocity dispersions for 10:1 mass ratio halos separated by the Lagrangian virial radius of the larger halo (dashed lines) as a function of halo mass (equation~(\protect\ref{eq:dispersion}) with the environmental factor of equation~(\protect\ref{eq:environment}) applied). Line colours correspond to different redshifts as indicated in the figure. In this figure velocities are shown in dimensionful units.}
\label{fig:velocities}
\end{figure}

Most strikingly, our model predicts that, at primary halo masses below those considered by \cite{li_orbital_2020}, the mean and dispersion in both radial and tangential velocities will increase significantly. This result can be understood by considering how the velocity dispersion of secondary halos varies as a function of primary halo mass. Figure~\ref{fig:velocities} compares halo virial velocities (shown in dimensionful units in this figure) with the linear theory velocity dispersion for 10:1 primary--secondary mass ratio halos separated by the Lagrangian virial radius of the primary halo (dashed lines) as a function of halo mass\footnote{We do not show comparisons to results from N-body simulations in this figure. The velocity dispersions shown are the linear theory expectation, computed at the Lagrangian virial radius of the primary halo (as this is the key quantity which goes into our model). What could be measured from N-body simulations is the velocity dispersion at the corresponding Eulerian separation but, by construction, that is far into the nonlinear regime, making a comparison to what is plotted here not meaningful. The model presented in this work takes the linear theory expectation for the velocity distribution, and propagates it into the deeply nonlinear regime. Therefore, the confrontation with N-body results in Figure~\protect\ref{fig:distributions} is the relevant comparison.}. At the highest masses shown the velocity dispersion of the secondary halos is small compared to the primary halo virial velocity. However, as halo mass decreases the virial velocity decreases faster than the velocity dispersion of the secondary halos. Consequently, for low mass primary halos the secondary halo velocity dispersion becomes comparable to, and can exceed, the primary halo virial velocity. This increased secondary halo velocity dispersion (when expressed relative to the primary halo virial velocity) is the cause of the increase in the mean and dispersion of $V_\mathrm{r}$ and $V_\theta$ for low mass primary halos. We find that the ratio $\sigma/V_\mathrm{v}$ is the primary driver of the shape of the orbital velocity distribution function.

\subsection{Redshift Dependence}

\begin{figure}
\includegraphics[width=85mm]{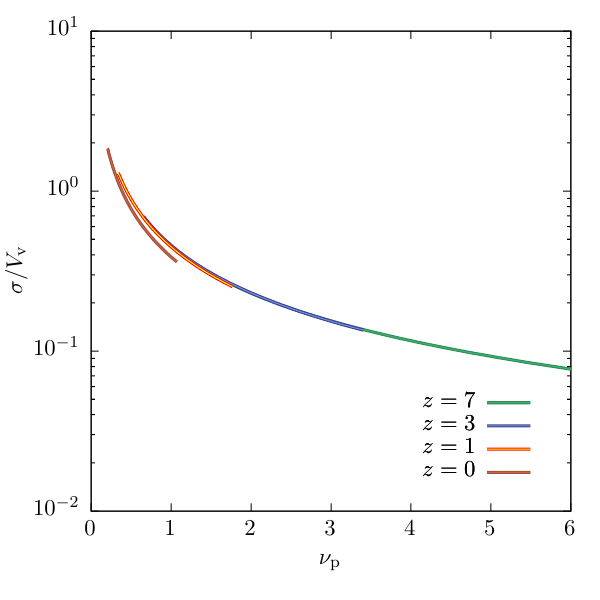}
\caption{Halo-halo pairwise velocity dispersions, in units of the primary halo virial velocity, for 10:1 mass ratio halos separated by the Lagrangian virial radius of the larger halo (dashed lines) as a function of peak height, $\nu_\mathrm{p}$. Line colours correspond to different redshifts as indicated in the figure.}
\label{fig:velocitiesPeakHeight}
\end{figure}

In Figure~\ref{fig:velocities} we also show results for different redshifts. At $z>0$ virial velocities are larger for fixed $M_\mathrm{p}$, while the velocity dispersion of the secondary halos is lower. \cite{li_orbital_2020} considered redshift dependence in the orbital velocity distribution, and showed that, when considered at fixed peak height, $\nu = \delta_\mathrm{c}/\sqrt{S}$, where $\delta_\mathrm{c}$ is the critical linear theory overdensity for halo collapse, there was little redshift dependence. We therefore recast Figure~\ref{fig:velocities} in terms of peak height, and plot the velocity dispersion of the secondary halos now in dimensionless units (i.e. in units of the primary halo virial velocity). The results are shown in Figure~\ref{fig:velocitiesPeakHeight}.

At fixed peak height, $\nu_\mathrm{p}$, there is little evolution in the background halo velocity dispersion in dimensionless units, and, therefore, little change in the predicted distribution of orbital velocities at fixed $\nu_\mathrm{p}$. To understand this behaviour we can consider the expected scaling of the relevant quantities with redshift. We characterize the present day matter power spectrum as a power law in wavenumber, $k$,
\begin{equation}
 P(k) \propto k^{n_\mathrm{eff}},
\end{equation}
where $n_\mathrm{eff}$ is the effective power law index on the scales of interest. Then, the root-variance of the density field, $\sqrt{S}(M_\mathrm{p},a)$, and the velocity dispersion of the background, $\sigma(M_\mathrm{p},a)$, as functions of the primary halo mass, $M_\mathrm{p}$, and expansion factor, $a$, are given by:
\begin{equation}
 \sqrt{S}(M_\mathrm{p},a) = D(a) \sigma_0(M_\mathrm{p}),
\end{equation}
and
\begin{equation}
 \sigma(M_\mathrm{p},a) = H(a) D(a) a f(a) \sigma_{-1}(M_\mathrm{p})
\end{equation}
respectively.

For our power law power spectrum we have (from equation~\ref{eq:sigmaj}):
\begin{equation}
 \sigma_j(M_\mathrm{p}) \propto k_\mathrm{p}^{(3+n_\mathrm{eff}+2j)/2} \propto M_\mathrm{p}^{-(3+n_\mathrm{eff}+2j)/6},
\end{equation}
where $k_\mathrm{p}$ is the wavenumber corresponding to the mass scale $M_\mathrm{p}$ in the window function $W(k|M)$.

The peak height is given by
\begin{equation}
 \nu_\mathrm{p} = \frac{\delta_\mathrm{c}(a)}{\sqrt{S}(M_\mathrm{p},a)} \propto \frac{\delta_\mathrm{c}(a)}{D(a)} M_\mathrm{p}^{(3+n_\mathrm{eff})/6},
\end{equation}
while the virial velocity is given by
\begin{equation}
 V_\mathrm{v} \propto \frac{M_\mathrm{p}}{r_\mathrm{v}} \propto M_\mathrm{p}^{1/3} a^{-1/2} \Delta_\mathrm{v}^{1/6}(a).
\end{equation}
Therefore
\begin{eqnarray}
 \frac{\sigma}{V_\mathrm{v}} &\propto& \frac{H(a) D(a) a f(a) M_\mathrm{p}^{-(1+n_\mathrm{eff})/6}}{M_\mathrm{p}^{1/3} a^{-1/2} \Delta_\mathrm{v}^{1/6}(a)} \nonumber \\
 &\propto& H(a) a^{3/2} D(a) f(a) \Delta_\mathrm{v}^{-1/6}(a) M_\mathrm{p}^{-(3+n_\mathrm{eff})/6}.
\end{eqnarray}
When expressed in terms of peak height this gives
\begin{equation}
  \frac{\sigma}{V_\mathrm{v}} \propto H(a) a^{3/2} f(a) \Delta_\mathrm{v}^{-1/6}(a) \delta_\mathrm{c}(a) \nu_\mathrm{p}^{-1}.
\end{equation}
Thus, the secondary halo velocity dispersion, expressed in virial units, is independent of $n_\mathrm{eff}$, and is expected to scale\footnote{This behaviour can be understood as follows. At fixed density, velocities in any self-gravitating system will scale as $V \propto r$. In cosmological linear theory the characteristic density is simply the mean density of the universe, independent of scale. For virialized dark matter halos the characteristic density is the virial density, also independent of scale. As such, any scale dependence is expected to cancel out in the ratio $\sigma/V_\mathrm{v}$. Additionally, $\sigma_\mathrm{V}^2 \propto S$ since both are derived from an integral over the power spectrum. Since $S \propto \nu_\mathrm{p}^{-2}$ this implied $\sigma_\mathrm{V} \propto \nu_\mathrm{p}^{-1}$.} as $\nu_\mathrm{p}^{-1}$, consistent with the behaviour seen in Figure~\ref{fig:velocitiesPeakHeight}. To examine the scaling with redshift, we first consider the case of an Einstein-de Sitter universe, which has the advantage of having simple scalings, and which approximates the actual universe at high redshift. For an Einstein-de Sitter universe $H(a) \propto a^{-3/2}$, while $f(a)$, $\Delta_\mathrm{v}(a)$, and $\delta_\mathrm{c}(a)$ are all independent of epoch, such that
\begin{equation}
  \frac{\sigma}{V_\mathrm{v}} \propto \nu_\mathrm{p}^{-1},
\end{equation}
i.e. there is no dependence on epoch. Our model therefore predicts, in an Einstein-de Sitter universe, that the distribution of orbital parameters (expressed in virial units) is independent of redshift when examined at fixed $\nu_\mathrm{p}$. In the actual cosmological model used in this work \citep{planck_collaboration_planck_2014}, we find that $\sigma/V_\mathrm{v}$ at fixed $\nu_\mathrm{p}$ is close to independent of redshift at $z > 1$ (where the Einstein-de Sitter approximation is reasonable), and decreases by around 16\% relative to its high-$z$ value by $z=0$. This is consistent with the behaviour seen in Figure~\ref{fig:velocitiesPeakHeight}.

\subsection{Truncated Power Spectra}

\begin{figure}
\includegraphics[width=85mm]{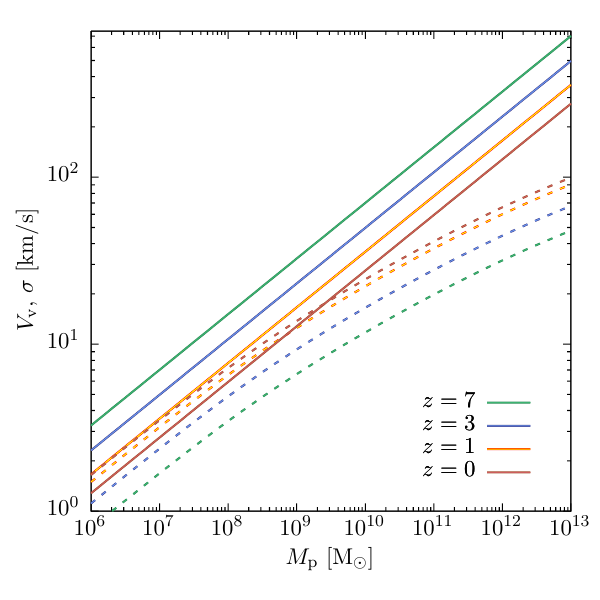}
\caption{Virial velocities (solid lines) and halo-halo pairwise velocity dispersions for 10:1 mass ratio halos separated by the virial radius of the larger halo (dashed lines) as a function of halo mass in a WDM model. Line colours correspond to different redshifts as indicated in the figure.}
\label{fig:velocitiesWDM}
\end{figure}

As the orbital velocity distribution in our model is controlled primarily by the ratio $\sigma/V_\mathrm{v}$ it will have some dependence on the matter power spectrum (which determines $\sigma$). In Figure~\ref{fig:velocitiesWDM} we examine how $\sigma$ is changed if we adopt a truncated power spectrum consistent with a 3~keV thermal warm dark matter particle. For such a particle the half-mode mass of the power spectrum ($M_\mathrm{hm}$; i.e. the mass corresponding to the wavenumber at which the transfer function is suppressed by a factor of 2 compared to the CDM case) is $M_\mathrm{hm}=4.1\times 10^8\mathrm{M}_\odot$ \citep{schneider_non-linear_2012}. Comparing with Figure~\ref{fig:velocities} it is clear that, while there is some reduction in $\sigma$ at low masses, it is quite small. In particular, at $z=0$ we find that $\sigma$ is reduced by around 30\% compared to the CDM case, even though this mass scale lies well below the half-mode mass. The reason for this is that, since the velocity field is sourced by the gradient of the density field, it gives a stronger weighting to large scale modes---in equation~(\ref{eq:sigmaj}) the power spectrum is weighted by $k^2$ when evaluating the variance of the density field, but by $k^0$ when evaluating the dispersion of the velocity field.

\begin{figure*}
\begin{tabular}{cc}
\includegraphics[width=85mm]{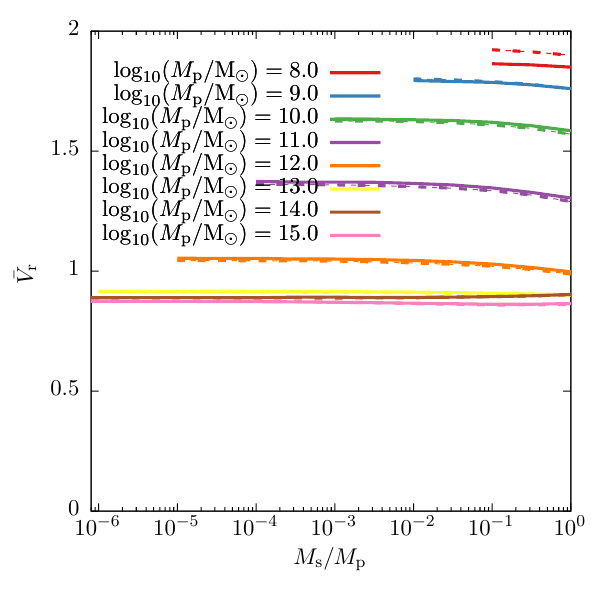} & \includegraphics[width=85mm]{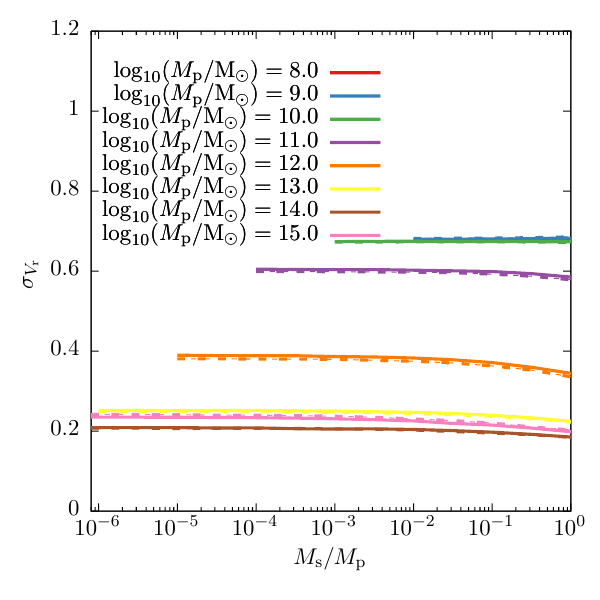}
\end{tabular}
 \caption{Summary statistics of the orbital velocity distribution function as predicted by our model for a thermal WDM power spectrum (solid lines), and for the equivalent CDM power spectrum (dashed lines). Statistics shown are the mean ($\bar{V}$; left panel), and dispersion ($\sigma$; right panel), of the radial orbital velocities, as a function of the secondary-to-primary halo mass ratio, $M_\mathrm{s}/M_\mathrm{p}$. Lines colours indicate different primary halo masses, as indicated in each panel.}
\label{fig:WDM}
\end{figure*}

Figure~\ref{fig:WDM} shows the mean and dispersion of the radial velocity distribution as a function of primary halo mass and the secondary-to-primary halo mass ratio for this WDM power spectrum (solid lines), and for the equivalent CDM power spectrum (dashed lines). There is almost no change in these summary statistics due to the truncation of the power spectrum in the WDM case, with the exception of a small reduction in the mean radial velocity for $M_\mathrm{p}=10^8\mathrm{M}_\odot$. Since a truncation of the power spectrum on these scales is already strongly ruled out---for example, \citealt{nadler_dark_2021} find that $m_\mathrm{WDM}>9.7~\hbox{keV}$ at 95\% confidence (corresponding to $M_\mathrm{hm}<10^{7.4}\mathrm{M}_\odot$)---we conclude that the effects of plausible changes in the power spectrum are unlikely to have a strong effect on the distribution of orbital velocities on mass scales of current astrophysical interest.

\section{Discussion}\label{sec:discussion}

We have described a simple model for the joint distribution function of the orbital velocities, $(V_\mathrm{r},V_\theta)$, of merging dark matter halo pairs. Our model considers a dark matter halo perturber, described by a Keplerian potential, embedded within a background of secondary halos with velocities as predicted by linear perturbation theory, and assuming a Gaussian distribution of pairwise velocities for the background halos\footnote{This assumption of Gaussianity could be relaxed, for example using the distribution functions found by \protect\cite{cuesta-lazaro_towards_2020}, although the improved precision in the characterization of the background halo velocity distribution may not be warranted given the other approximations made in our model.}. By considering secondary halos on orbits in the ``loss cone'' of the primary halo we derive the distribution of merging velocities. Despite its simplicity, and the fact that halo merging occurs far into the non-linear regime of structure formation, our model predictions are in excellent qualitative, and reasonable quantitative agreement with distributions measured from N-body simulations \citep{li_orbital_2020}. Our model, coupled with a description of the density profile of the inflowing stream of matter onto a primary halo, also produces good agreement with measurements of halo merger rates (see Appendix~\ref{sec:mergerRates}).

Assuming a potential corresponding to an extended profile for the primary halo (rather than a Keplerian potential) would lead to the velocity distribution to be shifted to higher velocity. For example, if we consider an Einasto profile (with concentration parameter, $c=10$, and shape parameter, $\alpha=0.18$, for specificity) then the potential difference from infinity to the virial radius is approximately $1.46 V_\mathrm{V}^2$ (as opposed to just $V_\mathrm{V}^2$ for the Keplerian potential) which would lead to velocities being around 20\% larger. However, the primary halo potential is growing via the accretion of secondary halos (and, perhaps, some smooth accretion of mass). In a simple model in which there is no crossing of mass shells during the growth of the primary \citep[e.g.][]{fillmore_self-similar_1984} then the mass contained within the current orbital position of an infalling secondary would be constant (as the secondary comoves with the overall mass shell), and the Keplerian potential approximation would be valid. This assumption will break down for at least two reasons: 1) once the secondary reaches the backsplash radius of the primary there \emph{is} shell crossing; and 2) since we are considering a distribution of secondary orbits, not all of them can comove with the mass shell.

We expect then that the Keplerian approximation is reasonable for the mean infall velocity (with some inaccuracy arising from shell-crossing close to the primary halo), but will be less accurate for halos in the tails of the distribution (which experience signficant shell-crossing during their infall). An improvement of our model could consider the assembly of the primary halo (perhaps using a secondary infall model; \citealt{fillmore_self-similar_1984,bertschinger_self-similar_1985}), and track the evolution of each secondary halo orbit in this evolving potential.

Our model connects the distribution of orbital velocities to the underlying matter power spectrum. This allows us to explore predictions for how the distribution of orbital parameters is expected to change for smaller primary halo masses, where we predict a significant increase in the width of these distributions, extending to higher velocities (when expressed in units of the characteristic virial velocity of the primary halo). Furthermore, this simple model shows that the quantity $\sigma/V_\mathrm{v}$---the ratio of the background velocity dispersion to the primary halo virial velocity---is the primary driver of the shape and width of the orbital velocity distribution. We show that, when considered at constant peak height, $\nu$, this quantity has a very weak dependence on redshift, in agreement with measurements from N-body simulations \citep{li_orbital_2020}.

The increased width of the merging halo velocity distributions for lower mass primary halos may have some important consequences. As these distributions set the initial conditions for further evolution of the subhalo within the potential of its host, they may affect the radial distribution of subhalos for example. However, primary halo mass scales on which measurements of the subhalo distribution are currently being made (of order $10^{12}\mathrm{M}_\odot$ for studies of dwarf galaxies around the Milky Way, and of order $10^{13}\mathrm{M}_\odot$ for studies of gravitational lensing around massive elliptical galaxies) are in the regime where $\sigma/V_\mathrm{v}$ is small, and, in any case, results from N-body simulations have been explored on these mass scales.

Other possible consequences of the increased width of the merging halo velocity distributions for lower mass primary halos arise from the connection between halo mergers and halo structure. For example, both halo spin \citep{vitvitska_origin_2002,benson_random-walk_2020} and concentration \citep{johnson_random_2021} appear to be connected to the orbital parameters of halo merger events. However, in cases where the merging halo velocity distribution width is increased many of the mergers will be unbound, and so may correspond to ``fly-by'' encounters which have little lasting impact on the primary halo. We intend to explore the consequences of our model on halo structure in a future work.

Further exploiting this direct connection to the power spectrum, we explore how the distribution of orbital velocities is changed in the case of a truncated power spectrum associated with the thermal warm dark matter model. We find that, given current constraints on the mass of a thermal warm dark matter particle, there are no significant changes in the distribution of orbital parameters on scales of current astrophysical interest. As such, it is reasonable to utilize fitting functions for the distribution of orbital velocities of merging dark matter halos that were derived from simulations of CDM even in the case of plausible WDM models.

In conclusion, our model provides physical insight into the origins of the distribution of orbital velocities of merging halo pairs, and a clear connection to the underlying matter power spectrum. While prior measurements of the orbital parameter distribution from N-body simulations have shown only a very weak dependence on halo mass, our model predicts a stronger dependence on mass scales smaller than those which have been explored in these prior studies. Our model therefore provides guidance on the extent of the parameter space over which fitting functions, often provided by those prior studies, can be considered to be valid. The code used to generate the distribution functions used in this work is publicly available as part of the \href{https://github.com/galacticusorg/galacticus}{\sc Galacticus} toolkit, which also allows those distribution functions to be utilized as part of other calculations made with the toolkit (e.g. evolving populations of subhalos).

\section*{Acknowledgements}

We thank Shaun Cole, Benedikt Diemer, Xiaolong Du, Fangzhou Jiang, and Annika Peter for helpful and insightful discussions. Computing resources used in this work were made available by a generous grant from the Ahmanson Foundation.

\section*{Data availability}

The data underlying this article will be shared on reasonable request to the corresponding author. The code used to generate the distribution functions used in this work is publicly available as part of the \href{https://github.com/galacticusorg/galacticus}{\sc Galacticus} toolkit.

\bibliographystyle{mn2e}
\bibliography{orbitalVelocities}

\appendix

\section{Orbital Velocity Distributions}\label{sec:appendix}

\begin{figure*}
 \begin{tabular}{cc}
  \includegraphics[width=85mm]{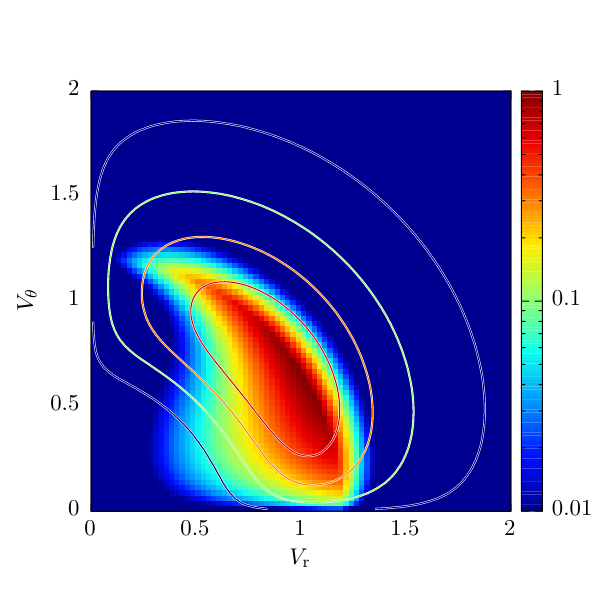} &
  \includegraphics[width=75mm]{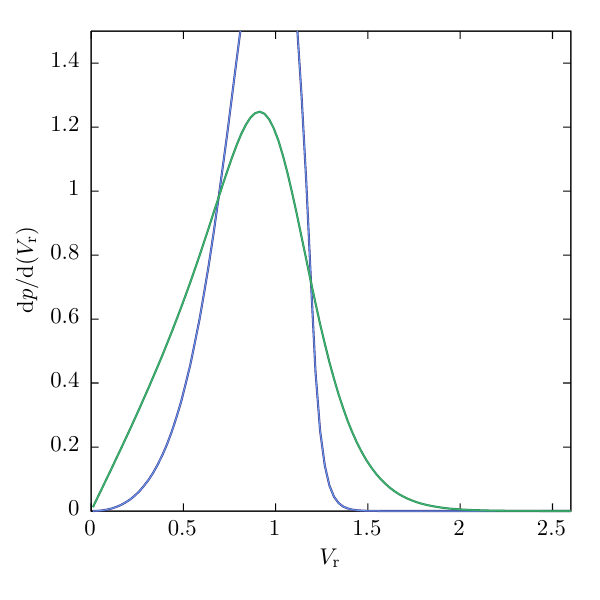} \\
  \includegraphics[width=75mm]{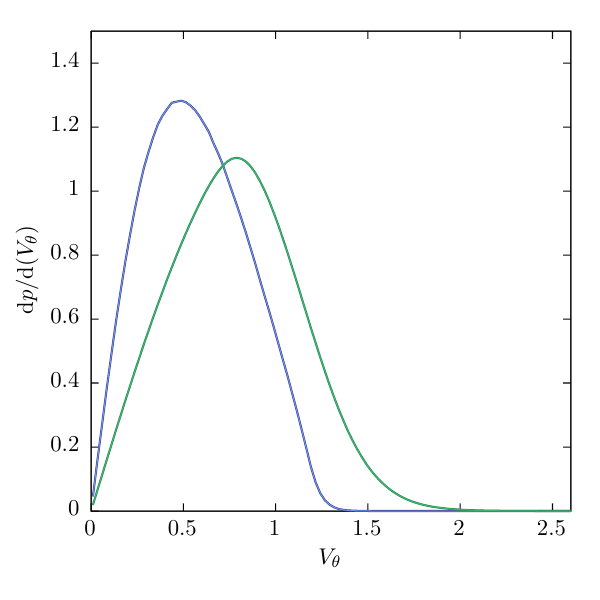} &
  \includegraphics[width=75mm]{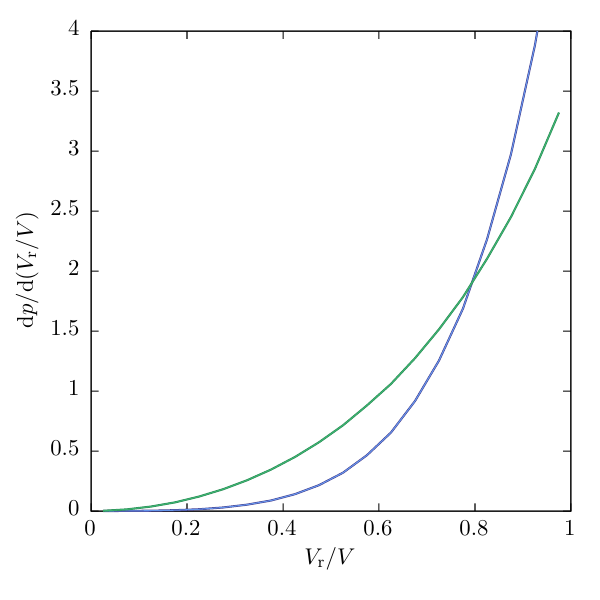}
 \end{tabular}
 \caption{Distributions of orbital parameters for merging halos of masses $M_\mathrm{p}=10^{14}\mathrm{M}_\odot$ and $M_\mathrm{s}=10^{11}\mathrm{M}_\odot$ at $z=0$. \emph{Upper left panel:} The joint distribution of radial and tangential velocities of merging halos. The colour map shows results from our model, while the contours indicate the distribution of \protect\cite{li_orbital_2020}. \emph{Upper right panel:} The distribution of radial velocities of merging halos. The blue curve shows results from our model, while the green line is the distribution of \protect\cite{li_orbital_2020}. \emph{Lower left panel:} As the upper right panel, but for the tangential velocity. \emph{Lower right panel:} As the upper right panel, but for the radial velocity expressed as a fraction of the total velocity.}
 \label{fig:distributions14}
\end{figure*}

Figure~\ref{fig:distributions14} shows distributions of orbital velocities as in Figure~\ref{fig:distributions} but for a much larger primary halo mass of $M_\mathrm{p}=10^{14}\mathrm{M}_\odot$. The secondary halo mass is unchanged from Figure~\ref{fig:distributions} at $M_\mathrm{s}=10^{11}\mathrm{M}_\odot$.

As expected from the summary statistics shown in Figure~\ref{fig:summaryStats} our model performs less well here compared to the results shown in Figure~\ref{fig:distributions}. Specifically, while the location of the peak in the radial velocity is well-matched to the results of \protect\cite{li_orbital_2020} the dispersion around this mean is too small in our model. This also leads to the distribution of tangential velocities being shifted toward low velocities compared to \protect\cite{li_orbital_2020}. In the top-left panel the joint distribution of $(v_\mathrm{r},v_\theta)$ retains the ridge-line of approximately constant orbital energy, but the distribution is too narrow around this locus.

Overall, this suggests that our model underestimates the ratio $\sigma/V_\mathrm{V}$ in this regime. Possible causes of this underestimate include the limitations of linear perturbation theory, and the approximate nature of our model for the environmental dependence of the secondary halo velocity dispersion.

\section{Halo Merger Rates}\label{sec:mergerRates}

\begin{figure}
  \includegraphics[width=75mm]{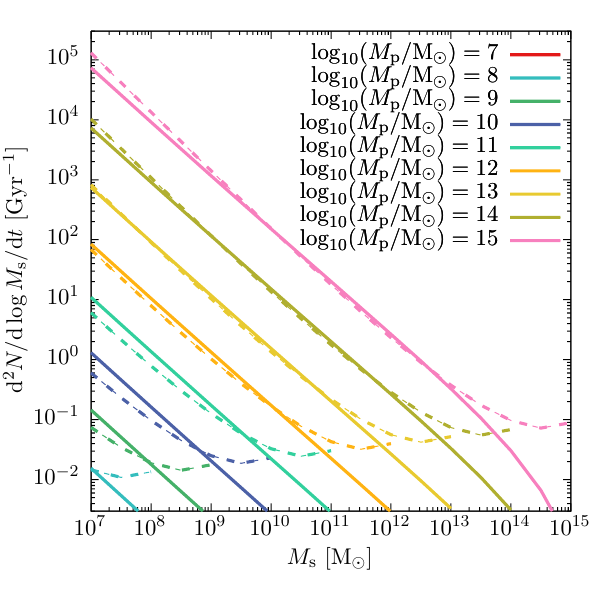} 
  \caption{The merger rate, at $z=0$, of secondary halos of mass $M_\mathrm{s}$ is shown for several different values of primary halo mass, $M_\mathrm{p}$ (as indicated by line colour). Dashed lines show merger rates measured by \protect\cite{fakhouri_merger_2010} while solid lines show estimates based on the results of this work.}
  \label{fig:mergerRates}
\end{figure}

Given knowledge of the mean radial velocity of merging halo pairs, along with an estimate of the number density of such pairs at the point of merging, it is straightforward to estimate the rate of mergers of primary-secondary halo pairs:
\begin{equation}
\frac{\mathrm{d}^2 N}{\mathrm{d}\log M_\mathrm{s} \mathrm{d} t} = 4 \pi r_\mathrm{V}^2 n(r_\mathrm{V},M_\mathrm{p},M_\mathrm{s}) \bar{v}_\mathrm{r}(M_\mathrm{p},M_\mathrm{s}),
\end{equation}
where $r_\mathrm{V}$ is the virial radius of the primary halo, and $n(r,M_\mathrm{p},M_\mathrm{s})$ is the number density of accreting secondary halos of mass $M_\mathrm{s}$ at separation $r$ from primary halos of mass $M_\mathrm{p}$.

Using our model, which predicts $\bar{v}_\mathrm{r}$, we can predict the expected merger rates of halos. For $n(r,M_\mathrm{p},M_\mathrm{s})$ we assume that the density profile of the inflow stream around the primary halo follows the form found by \protect\cite{diemer_dependence_2014}, and assume no bias between halos and mass in this stream such that $n(r,M_\mathrm{p},M_\mathrm{s}) = n(M_\mathrm{s}) \rho_\mathrm{i}(r)/\bar{\rho}$ where $n(M_\mathrm{s})$ is the usual halo mass function, $\rho_\mathrm{i}(r)$ is the density profile of the inflow stream from \protect\cite{diemer_dependence_2014}, and $\bar{\rho}$ is the mean density of the universe. The assumption of no bias in the inflow is unlikely to be true in detail but has not been extensively studied\footnote{Specifically, at the separations of concern here the inflow stream is spatially coincident with the backsplash halo population of the primary halo. Any determination of the bias of halos in the inflow stream would need to separate out these two populations.}.

The results of this simple model are shown, as a function of secondary halo mass and for several different primary halo masses, in Figure~\ref{fig:mergerRates}, where they are compared to the measured merger rates from \protect\cite{fakhouri_merger_2010}. At secondary-primary mass ratios less than around $0.1$ the agreement between the results of this simple model and the \protect\cite{fakhouri_merger_2010} measurements are very good---particularly for $10^{12}\mathrm{M}_\odot < M_\mathrm{p} < 10^{14}\mathrm{M}_\odot$, the regime where our model predictions for $\bar{v}_\mathrm{r}$ are in good agreement with N-body measurements. At lower values of $M_\mathrm{p}$ our model predicts merger rates slightly larger than those of \protect\cite{fakhouri_merger_2010}.

At higher mass ratios our model fails to capture the upturn in merger rates measured by \protect\cite{fakhouri_merger_2010}. While a detailed analysis of this discrepancy is beyond the scope of this work there are several plausible contributing factors. First, as can be seen in Figure~\ref{fig:summaryStats}, our model underestimates $\bar{v}_\mathrm{r}$ for large $M_\mathrm{s}/M_\mathrm{p}$, which would lead to an underestimate of the merger rate. Perhaps more importantly the perturbative nature of our model (which assumes that the primary halo is a perturber in the background of secondary halos) must break down when the secondary halo mass becomes comparable to the primary halo mass. Finally, as mentioned above, our assumption of no bias for the density profile of secondary halos in the inflow stream is likely to be incorrect and may contribute to the discrepancy at large $M_\mathrm{s}/M_\mathrm{p}$.

Using these estimates of halo merger rates the model described in this work could be used to compute the mass growth rates of primary halos. To do so would require assessing which merging secondaries are actually bound to the primary. \protect\cite{benson_orbital_2005} showed that around 18\% of merging secondaries are on unbound orbits, but that the majority of these become bound (either due to the effects of dynamical friction or because the primary halo continues to grow after the secondary becomes a subhalo within it). Therefore, to a reasonable approximation all merging secondary halos contribute to the long-term mass growth of the primary. However, the results of \protect\cite{benson_orbital_2005} apply to halos of masses of around $10^{11}\mathrm{M}_\odot$ and greater, and so may not be valid for lower mass halos where ths model described in this work predicts a much larger fraction of unbound orbits. Computing primary halo growth rates using this approach will therefore require a careful treatment of the orbital evolution of each subhalo and the growth of the primary halo potential.

\end{document}